\documentclass[prd,preprint,superscriptaddress,showpacs,byrevtex]{revtex4}
\usepackage{bm}
\def\fsl#1{\setbox0=\hbox{$#1$}           
   \dimen0=\wd0                                 
   \setbox1=\hbox{/} \dimen1=\wd1               
   \ifdim\dimen0>\dimen1                        
      \rlap{\hbox to \dimen0{\hfil/\hfil}}      
      #1                                        
   \else                                        
      \rlap{\hbox to \dimen1{\hfil$#1$\hfil}}   
      /                                         
   \fi}                                         %
\newcommand{\be}{\begin{equation}}
\newcommand{\ee}{\end{equation}}
\newcommand{\bea}{\begin{eqnarray}}
\newcommand{\eea}{\end{eqnarray}}
\newcommand{\beq}{\begin{equation}}
\newcommand{\eeq}{\end{equation}}
\newcommand{\beqs}{\begin{eqnarray}}
\newcommand{\eeqs}{\end{eqnarray}}

\begin{document}
\title{ Angular Momentum Sum Rule Violation in QCD Due to Confinement and The Proton Spin Crisis }
\author{Gouranga C Nayak }\thanks{G. C. Nayak was affiliated with C. N. Yang Institute for Theoretical Physics in 2004-2007.}
\affiliation{ C. N. Yang Institute for Theoretical Physics, Stony Brook University, Stony Brook NY, 11794-3840 USA}
\date{\today}
\begin{abstract}
The proton spin crisis remains an unsolved problem in physics. In this paper we find that due to the confinement of partons inside the hadron the angular momentum sum rule in QCD is violated. Hence we find that the non-vanishing angular momentum flux contribution of the partons in QCD should be added to the spin and angular momentum of the partons to solve the proton spin crisis.
\end{abstract}
\maketitle
\pagestyle{plain}

\pagenumbering{arabic}

\section{Introduction}

A proton in motion at the high energy colliders consists of quarks, antiquarks and gluons. This is because the sea quark-antiquark pairs and gluons are produced from the QCD vacuum via vacuum polarization at very high energy. In the parton model description of the proton at high energy colliders it was predicted that the spin of the proton is the sum of the spin of the partons inside the proton \cite{jaff}.

However, the experimental data suggested otherwise. In fact in the very early experimental measurement by EMC in the late 1980's \cite{emcc} it was found that the sum of the spin of quarks plus antiquarks is nearly nil. For the updates on world data on proton spin crisis, see \cite{oher}.

Recently the direct measurement of the spin dependent gluon distribution function inside the proton is reported by the RHIC experiment \cite{hc}. In addition to the direct measurement of the spin dependent gluon distribution function, the spin physics experiment at RHIC also provides an opportunity to test spin transfer processes in pQCD \cite{nsm}. To summarize the overall experimental status, the world data on spin dependent parton distribution functions predicts that about fifty percent \cite{fifty} of the proton spin is carried by the spin of the partons inside the proton.

Since about fifty percent of the proton spin is missing we have the proton spin crisis. In order to resolve this crisis it is suggested that the orbital angular momentum of the partons should be added to the spin angular momentum of the partons \cite{jaff}.

This, however, is not easy because of the issue of gauge invariance in QCD. In fact one can not define gauge invariant spin or orbital angular momentum of the Yang-Mills field in the Yang-Mills theory \cite{nmx,nym}. The total angular momentum of the Yang-Mills field is gauge invariant but the total angular momentum of the Yang-Mills field is not the sum of the spin and orbital angular momentum of the Yang-Mills field due to non-vanishing boundary surface term in the Yang-Mills theory \cite{nmx,nym}.

These issues of the gauge invariance suggests that the parton model prediction \cite{jaff,all1} should read as
\bea
&&\frac{1}{2} =\sum_{  q}<{\hat { S}}_q>+\sum_{  q} <{\hat { L}}_q>+\sum_{\bar  q} <{\hat { S}}_{\bar q}>+\sum_{\bar  q} <{\hat { L}}_{\bar q}>+\sum_{g}<{\hat { J}}_g>+<{\hat { J}}_{\rm flux}>\nonumber \\
&& \neq \sum_{  q}< {\hat  S}_q >+\sum_{ q}< {\hat {\tilde L}}_q >+\sum_{\bar  q}<{\hat  S}_{\bar q} >+\sum_{\bar  q}< {\hat {\tilde L}}_{\bar q} >+\sum_{g}< {\hat  {\tilde S}}_g >+\sum_{g}< {\hat  {\tilde L}}_g >
\label{neqs}
\eea
where $\frac{1}{2}$ in the left hand side is the spin of the proton and ${\hat {\tilde S}}_g$, ${\hat {\tilde L}}_g$ are the gauge non-invariant spin, orbital angular momentum operators of gluon, ${\hat { J}}_g$ is the gauge invariant total angular momentum operator of gluon, ${\hat {\tilde L}}_{q({\bar q})}$ is the gauge non-invariant orbital angular momentum operator of quark  (antiquark) and ${\hat S}_{q({\bar q})}$, ${\hat L}_{q({\bar q})}$ are the gauge invariant spin, orbital angular momentum operators of quark (antiquark) \cite{nmx,nym}. In eq. (\ref{neqs})
\bea
< {\hat O}>=<p,s_z|{\hat O}|p,s>
\label{expt}
\eea
where ${\hat O}$ is the longitudinal component of the vector operator ${\hat {\vec O}}$ and $|p,s>$ is the longitudinally polarized proton (physical) state with momentum $p$ and spin $s$. In eq. (\ref{neqs}) the ${\sum}_{q,{\bar q},g}$ includes all the quarks, antiquarks, gluons inside the proton.

In this paper we show that eq. (\ref{neqs}) is also not true due to confinement, {\it i. e.},
\bea
&&\frac{1}{2} \neq \sum_{  q}<{\hat { S}}_q>+\sum_{  q} <{\hat { L}}_q>+\sum_{\bar  q} <{\hat { S}}_{\bar q}>+\sum_{\bar  q} <{\hat { L}}_{\bar q}>+\sum_{g}<{\hat { J}}_g>\nonumber \\
&& \neq \sum_{ q}< {\hat  S}_q >+\sum_{ q}< {\hat {\tilde L}}_q >+\sum_{\bar  q}<{\hat  S}_{\bar q} >+\sum_{\bar  q}< {\hat {\tilde L}}_{\bar q} >+\sum_{g}< {\hat  {\tilde S}}_g >+\sum_{g}< {\hat  {\tilde L}}_g >.
\label{ini2}
\eea
In this paper we find that
\bea
\frac{1}{2}=\sum_{  q}<{\hat { S}}_q>+\sum_{  q} <{\hat { L}}_q>+\sum_{\bar  q} <{\hat { S}}_{\bar q}>+\sum_{\bar  q} <{\hat { L}}_{\bar q}>+\sum_{g}<{\hat { J}}_g>+<{\hat { J}}_{\rm flux}>
\label{ini1}
\eea
where ${\hat { J}}_{\rm flux}$ is the gauge invariant angular momentum flux operator of all the partons inside the proton in QCD which is non-vanishing due to the confinement of partons inside the proton. The definition of the gauge invariant angular momentum flux ${ J}_{\rm flux}$ in the Yang-Mills theory is given in eq. (\ref{jfx}). The eq. (\ref{ini2}) follows from eq. (\ref{ini1}).

Note that the concept of the angular momentum flux in physics is not new, see for example \cite{barn}, for a study of the non-vanishing angular momentum flux of the light.

Hence in this paper we find that due to the confinement of partons inside the hadron there exists non-vanishing angular momentum flux in QCD. Therefore we predict that the non-vanishing angular momentum flux contribution of the partons in QCD should be added to the spin and angular momentum of the partons to solve the proton spin crisis.

The paper is organized as follows. In section II we briefly discuss the vanishing angular momentum flux in the Dirac-Maxwell theory. In section III we discuss the non-perturbative QCD at infinity and the classical Yang-Mills theory. In section IV we show that due to the confinement of partons there exists non-vanishing angular momentum flux in the Yang-Mills theory. In section V we discuss the non-vanishing angular momentum flux contribution of partons to the proton spin in QCD. In section VI we show that the absence of non-vanishing boundary surface term in the study of the proton spin crisis in the literature is not justified. In section VII we show that angular momentum flux at infinite distance is not same as the anomalous glue contribution at Bjorken $x=0$. Section VIII contains conclusions.

\section{ Vanishing Angular Momentum Flux in Dirac-Maxwell Theory }

The conservation equation of the gauge invariant angular momentum obtained from the first principle in the Dirac-Maxwell theory gives \cite{nmx}
\bea
&&\frac{dJ^k_{EM}}{dt}+\frac{dJ^k_e}{dt}=-\epsilon^{knj} \int d^3x ~\partial_l [ x^n \{E^j(x) E^{l}(x)\nonumber \\
&&-\delta^{lj}\frac{{\vec E}^2(x) -{\vec B}^2(x)}{2}+\frac{i}{2} {\bar \psi}(x)[\gamma^l  ({\overrightarrow \partial}^j +ieA^j(x)) -\gamma^l ({\overleftarrow \partial}^j -ieA^j(x)) ] \psi(x)\} \nonumber \\
&&+ \frac{1}{8}{\bar \psi}(x) \{\gamma^l,~\sigma^{nj} \} \psi(x)]
\label{dace4}
\eea
where $A^\mu(x)$ is the electromagnetic potential, ${\vec E}$ (${\vec B}$) is the electric (magnetic) field, $\psi(x)$ is the Dirac field of the electron and $J^k_e$ ($J^k_{EM}$) is the gauge invariant total angular momentum of the electron (the electromagnetic field) given by
\bea
&&{ J}^k_e = \int d^3x~ [\epsilon^{klj} x^l \times [\frac{1}{2}\psi^\dagger(x) [-i{\overrightarrow { D}^j}+i{\overleftarrow {D}^j}] \psi(x)] +\psi^\dagger(x) ~{\Sigma}^k~ \psi(x)],\nonumber \\
&&{J}^k_{\rm EM}=\int d^3r ~ \epsilon^{klj} x^l ~ [{\vec E}(x) \times {\vec B}(x)]^j,~~~~~~{ D}^k=  { \partial}^k +ie{ A}^k(x).
\label{teman}
\eea

Since the electromagnetic potential $A_\mu(t,r)$ produced by the electron falls off as $\frac{1}{r}$ we find after integrating over the infinite volume the vanishing boundary surface term
\bea
&&  \int d^3x ~\partial_l [ x^n \{E^j(x) E^{l}(x)-\delta^{lj}\frac{{\vec E}^2(x) -{\vec B}^2(x)}{2}\nonumber \\
&&+\frac{i}{2} {\bar \psi}(x)[\gamma^l  ({\overrightarrow \partial}^j +ieA^j(x)) -\gamma^l ({\overleftarrow \partial}^j -ieA^j(x)) ] \psi(x)\} + \frac{1}{8}{\bar \psi}(x) \{\gamma^l,~\sigma^{nj} \} \psi(x)]=0.
\label{bce4}
\eea

Using eq. (\ref{bce4}) in (\ref{dace4}) we find
\bea
\frac{d[ {\vec J}_e+{\vec J}_{\rm EM}]}{dt}=0
\label{acef}
\eea
which proves that the angular momentum of the electron plus the electromagnetic field is conserved in the Dirac-Maxwell theory when infinite volume is considered.

\section{ Non-Perturbative QCD at Infinity and The Classical Yang-Mills Theory}\label{nqcd}

The quantum electrodynamics (QED) and quantum chromodynamics (QCD) are two fundamental theories of the nature describing the physics of the electromagnetic force and the strong force respectively. The QED is the quantum field theory of the classical electrodynamics and QCD is the quantum field theory of the classical Yang-Mills theory. The perturbative QED (pQED) and the perturbative QCD (pQCD) calculations are useful when the QED coupling constant and the QCD coupling constant become small. The QED coupling constant becomes small (large) at long (short) distance whereas the QCD coupling constant becomes small (large) at short (long) distance. Hence the QED becomes non-perturbative at short distance and the QCD becomes non-perturbative at long distance.

Note that while the pQED and pQCD calculations are known up to certain orders in the coupling constant, but the non-perturbative QED and the non-perturbative calculations are not well known. In particular the analytical study of the non-perturbative QCD becomes difficult because the gluons are self interacting (a feature absent in QED) and the QCD coupling constant is large at long distance. In order to know how the hadron is formed from the quarks and/or antiquarks and gluons in QCD one needs to solve non-perturbative QCD. Although the lattice QCD uses numerical method to study non-perturbative QCD but the analytical study of the non-perturbative QCD is lacking.

Irrespective of this limitation in our understanding of the analytical study of the non-perturbative QCD we can still predict if the boundary surface term at infinite distance in QCD is zero or non-zero. This can be shown as follows.

First of all note that the $\frac{1}{r}$ form of the Coulomb potential produced by the electron can be found by solving the classical Maxwell equation. Similarly the $\frac{1}{r}$ form of the Coulomb potential can be found from the QED at long distance. At the short distance the QED potential becomes different from the Coulomb potential due to the quantum loop effects (vacuum polarization effects). However, at long distance the QED predicts the $\frac{1}{r}$ form of the Coulomb potential. Although one can calculate the quantum loop effects (vacuum polarization effects) in QED at long distance but these quantum loop effects (vacuum polarization effects) are zero at infinite distance where the boundary surface term occurs. Hence as far as the boundary surface term at infinity is concerned the QED predicts the $\frac{1}{r}$ Coulomb potential which is the same potential predicted in the classical Maxwell theory.

Hence we find that if the boundary surface term at infinity is zero in the classical Maxwell theory then the corresponding boundary surface term at infinity is also zero in QED. We can use this analogy between the QED at infinite distance and the classical Maxwell theory to the QCD at infinite distance and the classical Yang-Mills theory. This is because the Yang-Mills theory was discovered by making analogy with the Maxwell theory by extending the U(1) gauge group to the SU(3) gauge group \cite{cny}.

For example, in analogy to the covariant derivative $D_\nu=\partial_\nu -ieA_\nu(x)$ in the U(1) gauge theory (the Maxwell theory) the covariant derivative in the Yang-Mills theory is taken to be $D_\nu =\partial_\nu +igT^dA_\nu^d(x)$ by extending U(1) gauge group to SU(3) gauge group \cite{cny}. Similarly in analogy to the form of the Maxwell field tensor $F_{\nu \delta}(x)$ which is obtained from the commutation relation $[D_\nu,D_\delta]=-ieF_{\nu \delta}(x)$ in the U(1) gauge theory (the Maxwell theory) the form of the Yang-Mills field tensor $F_{\nu \delta}^d(x)$ is taken from the commutation relation $[D_\nu,D_\delta]=igT^dF_{\nu \delta}^d(x)$ in the Yang-Mills theory theory by extending U(1) gauge group to SU(3) gauge group \cite{cny}. Similarly in analogy to the form of the Maxwell field lagrangian density $-\frac{1}{4}F_{\nu \delta}(x)F^{\nu \delta}(x)$ in the U(1) gauge theory (the Maxwell theory) the form of the Yang-Mills field lagrangian density is taken to be $-\frac{1}{4}F_{\nu \delta}(x)F^{\nu \delta}(x)$ in the Yang-Mills theory by extending U(1) gauge group to SU(3) gauge group \cite{cny}. Similarly in analogy to the form of the Maxwell potential (electromagnetic potential) $A_\nu(x)$ which is obtained from the U(1) pure gauge potential $A^\nu_{pure}(x)$, the form of the Yang-Mills potential (color potential) $A_\nu^d(x)$ is obtained from the SU(3) pure gauge potential $A^{\nu d}_{pure}(x)$ in the Yang-Mills theory by extending U(1) gauge group to SU(3) gauge group \cite{cpn,ccn}.

Hence we find that the Yang-Mills theory is obtained by making analogy with the Maxwell theory by extending the U(1) gauge group to the SU(3) gauge group \cite{cny,cpn,ccn}. From the earlier discussion in this section we saw that as far as the boundary surface term at infinity is concerned the QED predicts the Coulomb potential which is the same potential that is predicted by the classical Maxwell theory. This implies that if the boundary surface term at infinity is zero in the classical Maxwell theory then the corresponding boundary surface term at infinity is zero in QED. Similarly since the Yang-Mills theory is obtained by making analogy with the Maxwell theory by extending the U(1) gauge group to the SU(3) gauge group \cite{cny,cpn,ccn} we find that as far as the boundary surface term at infinity is concerned the QCD predicts the same potential that is predicted by the classical Yang-Mills theory. This implies that if the boundary surface term at infinity is non-zero in the classical Yang-Mills theory then the corresponding boundary surface term at infinity is non-zero in QCD.

\section{ Confinement and Non-Vanishing Angular Momentum Flux in Yang-Mills Theory  }\label{yNoe}

Since we have not experimentally observed isolated quarks and/or antiquarks we find that the quarks and/or antiquarks are confined inside the hadron. QCD is the fundamental theory of the nature describing the interaction between quarks and gluons. Hence one of the central goal in the fundamental physics of the nature is to describe the formation of hadron from quarks and/or antiquarks and gluons using QCD. Since QCD is the quantum field theory of the classical Yang-Mills theory let us consider a system consisting of quarks and antiquarks and the Yang-Mills potential (color potential) $A_\lambda^d(x)$ in the classical Yang-Mills theory before considering QCD.

For a system consisting of quarks and antiquarks and the Yang-Mills potential (color potential) $A_\lambda^d(x)$ we find from the gauge invariant Noether's theorem in the classical Yang-Mills theory \cite{nym}
\bea
&&\frac{dJ^k_{YM}}{dt}+\sum_{ q} \frac{dJ^k_q}{dt}+ \sum_{\bar  q}\frac{dJ^k_{\bar q}}{dt}=-\epsilon^{kpj} \int d^3x~ \partial_l [ x^p [E^{jd}(x) E^{ld}(x)-\delta^{lj}\frac{
{\vec E}^d(x) \cdot {\vec E}^d(x)-{\vec B}^d(x) \cdot {\vec B}^d(x)}{2}]\nonumber \\
&&+\sum_{ q} [x^p\frac{i}{2} {\bar \psi}_n(x)[\gamma^l  (\delta_{nn'}{\overrightarrow \partial}^j -igT^d_{nn'}A^{j d}(x)) -\gamma^l ({\overleftarrow \partial}^j \delta_{nn'}+igT^d_{nn'}A^{j d}(x))
 ] \psi_{n'}(x) \nonumber \\
&&+ \frac{1}{8} \sum_{  q}{\bar \psi}_n(x) \{\gamma^l,~\sigma^{p j } \} \psi_n(x)] +({ antiquarks})~]
\label{ace4}
\eea
where ${\vec E}^d$ (${\vec B}^d$) is the chromo-electric (chromo-magnetic) field produced by the quarks plus antiquarks in the system, $A_\nu^d(x)$ is the color potential (Yang-Mills potential) produced by the quarks plus antiquarks in the system, ${\vec J}_{YM}$ (${\vec J}_q)$ is the gauge invariant total angular momentum of the Yang-Mills field (quark) given by
\bea
&& {J}^k_{YM}=\int d^3x ~ \epsilon^{kpj} ~x^p~  [{\vec E}^d(t,{\bf x}) \times {\vec B}^d(t,{\bf x})]^j \nonumber \\
&&{ J}^k_q =\int d^3x~ [ \epsilon^{kpj}~x^p~ \times [\frac{1}{2}\psi^\dagger_n(x) [-i{\overrightarrow { D}}^j_{nl}+i{\overleftarrow { D}}^j_{nl}] \psi_l(x)] +\psi^\dagger_l(x) ~{\Sigma}^k~ \psi_l(x)].
\label{aym}
\eea
In the above $D^j_{kl}[A]=\delta_{kl}\partial^j -igT^d_{kl}A^{jd}(x)$, $({ antiquarks})$ represents the corresponding boundary surface term for antiquarks, $\sum_q$ represents sum over all the quarks in the system and ${\bar q}$ represents antiquark. As mentioned above since isolated quarks and/or antiquarks are not experimentally observed but the quarks and/or antiquarks are confined inside the hadron, whenever we say the color field produced by the quarks and/or antquarks in this paper we mean the color field inside the hadron.

Note that the boundary surface term in eq. (\ref{ace4}) is non-zero if the boundary surface is at the finite distance. Hence we need to find out whether the boundary surface term in eq. (\ref{ace4}) is zero or non-zero when the boundary surface is at infinite distance.

For confinement to happen one finds that the chromo-electric field energy density $\frac{{\vec E}^d(t,r)\cdot {\vec E}^d(t,r)}{2}$ produced by the quarks and/or antiquarks can not fall faster than $\frac{1}{r^3}$ at long distance which means the chromo-electric field ${\vec E}^d(t,r)$ produced by the quarks and/or antiquarks can not fall faster than $\frac{1}{r^{\frac{3}{2}}}$. Note that this form of the chromo-electric field ${\vec E}^d(t,r)$ produced by the quarks and/or antiquarks is consistent with the form of the chromo-electric field ${\vec E}^d(t,r)$ produced by a single quark in the classical Yang-Mills theory which is not of the form $\frac{1}{r^2}$ \cite{cpn}. This is because if the chromo-electric field ${\vec E}^d(t,r)$ produced by a single quark in the classical Yang-Mills theory becomes $\frac{1}{r^2}$ form then the color charge of the quark becomes constant \cite{cpn,ccn} in which case the chromo-electric field ${\vec E}^d(t,r)$ produced by the quarks and/or antiquarks becomes $\frac{1}{r^2}$ form [similar to the Maxwell theory] which can not explain confinement. Hence one finds that because of the confinement, the chromo-electric field ${\vec E}^d(t,r)$ produced by the quarks plus antiquarks can not fall faster than $\frac{1}{r^{\frac{3}{2}}}$ which means non-vanishing boundary surface term
\bea
&&\int d^3x~ \partial_l [ x^p [E^{jd}(x) E^{ld}(x)-\delta^{lj}\frac{
{\vec E}^d(x) \cdot {\vec E}^d(x)-{\vec B}^d(x) \cdot {\vec B}^d(x)}{2}]\nonumber \\
&&+\sum_{ q} [x^p\frac{i}{2} {\bar \psi}_n(x)[\gamma^l  (\delta_{nn'}{\overrightarrow \partial}^j -igT^d_{nn'}A^{j d}(x)) -\gamma^l ({\overleftarrow \partial}^j \delta_{nn'}+igT^d_{nn'}A^{j d}(x))
 ] \psi_{n'}(x) \nonumber \\
&&+ \frac{1}{8} \sum_{  q}{\bar \psi}_n(x) \{\gamma^l,~\sigma^{p j } \} \psi_n(x)] +({ antiquarks})~]\neq 0.
\label{cn1}
\eea
Note that the non-vanishing boundary surface term in eq. (\ref{cn1}) is different from the non-vanishing boundary surface terms in \cite{nmx,nym,nconf,ldn}.

From eqs. (\ref{cn1}) and (\ref{ace4}) we find in the classical Yang-Mills theory
\bea
\sum_{ q}\frac{d {\vec J}_q}{dt}+\sum_{\bar  q}\frac{d {\vec J}_{\bar q}}{dt}+\frac{d{\vec J}_{YM}}{dt}+\frac{d{\vec J}_{\rm flux}}{dt}= 0
\label{fne1}
\eea
where the non-vanishing gauge invariant angular momentum flux ${\vec J}_{\rm flux}$ of all the quarks plus antiquarks plus the color field (Yang-Mills field) in the system in the classical Yang-Mills theory is given by
\bea
&&J^k_{\rm flux}=\epsilon^{kpj} \int d^4x~ \partial_l [ x^p [E^{jd}(x) E^{ld}(x)-\delta^{lj}\frac{
{\vec E}^d(x) \cdot {\vec E}^d(x)-{\vec B}^d(x) \cdot {\vec B}^d(x)}{2}]\nonumber \\
&&+\sum_{ q} [x^p\frac{i}{2} {\bar \psi}_n(x)[\gamma^l  (\delta_{nn'}{\overrightarrow \partial}^j -igT^d_{nn'}A^{j d}(x)) -\gamma^l ({\overleftarrow \partial}^j \delta_{nn'}+igT^d_{nn'}A^{j d}(x))
 ] \psi_{n'}(x) \nonumber \\
&&+ \frac{1}{8} \sum_{\bar  q}{\bar \psi}_n(x) \{\gamma^l,~\sigma^{p j } \} \psi_n(x)] +({ antiquarks})~]\neq 0.
\label{jfx}
\eea
In eq. (\ref{jfx}) the time integral $\int dt$ is indefinite integral but the volume integral $\int d^3x$ is definite integral where $\int d^4x=\int dt~\int d^3x$.

Unlike the gauge invariant total angular momentum ${\vec J}_{YM}$ of the Yang-Mills field which can not be split into spin plus orbital angular momentum parts due to non-vanishing boundary surface term \cite{nmx,nym} we can write for quark
\bea
&&{\vec J}_q = {\vec L}_q +{\vec S}_q,~~~~~~~~~~~{ S}^k_q = \int d^3x~ \psi^\dagger_l(x) ~{\Sigma}^k~ \psi_l(x), \nonumber \\
&&{ L}^k_q = \int d^3x~ \epsilon^{kpj}~x^p~ \times [\frac{1}{2}\psi^\dagger_n(x) [-i{\overrightarrow { D}}^j_{nl}+i{\overleftarrow { D}}^j_{nl}] \psi_l(x)].
\label{ace5x}
\eea
where ${\vec L}_q$ (${\vec S}_q$) is the gauge invariant orbital (spin) angular momentum of the quark. Using eq. (\ref{ace5x}) in (\ref{fne1}) we find
\bea
\sum_{ q}\frac{d {\vec S}_q}{dt}+\sum_{  q}\frac{d {\vec L}_q}{dt}+\sum_{\bar  q}\frac{d {\vec S}_{\bar q}}{dt}+\sum_{\bar  q}\frac{d {\vec L}_{\bar q}}{dt}+\frac{d{\vec J}_{YM}}{dt}+\frac{d{\vec J}_{\rm flux}}{dt}= 0
\label{fne2}
\eea
which means the total conserved angular momentum ${\vec J}_{total}$ of the system consisting of quarks plus antiquarks plus the Yang-Mills field is given by
\bea
{\vec J}_{total}=\sum_{  q}{\vec S}_q+\sum_{  q}{\vec L}_q+\sum_{\bar  q}{\vec S}_{\bar q}+\sum_{\bar  q}{\vec L}_{\bar q}+{\vec J}_{YM}+{\vec J}_{\rm flux}.
\label{jtot}
\eea

In the literature the total conserved angular momentum in the classical Yang-Mills theory is assumed to be ${\vec S}_q+{\vec L}_q+{\vec S}_{\bar q}+{\vec L}_{\bar q}+{\vec J}_{YM}$ which is valid if the boundary surface term vanishes. However, as shown in eq. (\ref{cn1}) the boundary surface term is non-zero in the classical Yang-Mills theory which gives non-vanishing angular momentum flux $J_{\rm flux}$, see eq. (\ref{jfx}).  Hence we find that the non-vanishing angular momentum flux $J_{\rm flux}$ should be added to ${\vec S}_q+{\vec L}_q+{\vec S}_{\bar q}+{\vec L}_{\bar q}+{\vec J}_{YM}$ to obtain the total conserved angular momentum ${\vec J}_{total}$ of the system in the classical Yang-Mills theory, see eq. (\ref{jtot}).

\section{ Angular Momentum Flux Contribution To The Proton Spin in QCD }\label{qcd}

In the classical Yang-Mills theory the fields are c-numbers but in QCD the fields are operators. Hence extending eq. (\ref{fne2}) from the classical Yang-Mills theory to QCD to study the proton spin we find
\bea
\sum_{ q}\frac{d <{\hat {\vec S}}_q>}{dt}+\sum_{  q}\frac{d <{\hat {\vec L}}_q>}{dt}+\sum_{\bar  q}\frac{d <{\hat {\vec S}}_{\bar q}>}{dt}+\sum_{\bar  q}\frac{d <{\hat {\vec L}}_{\bar q}>}{dt}+\sum_{ g}\frac{d<{\hat {\vec J}}_g>}{dt}+\frac{d<{\hat {\vec J}}_{\rm flux}>}{dt}= 0\nonumber \\
\label{fne2x}
\eea
where the hat means operators corresponding to respective terms in eq. (\ref{fne2}) which are obtained by replacing $\psi,~{\bar \psi},~A \rightarrow {\hat \psi},~{\hat {\bar \psi}},~{\hat Q}$ and the expectation $<...>$ is defined in eq. (\ref{expt}), the symbol $g$ stands for gluon and $\sum_g$ means sum over all the gluons (hard and soft) present inside the proton. Note that we have used the notation ${\hat Q}$ for gluon field instead of ${\hat A}$ as we use the notation $A$ for the background field in the background field method of QCD which is used to prove factorization of infrared (IR) and collinear divergences and renormalization of ultra violet (UV) divergences in QCD at high energy colliders at all orders of coupling constant \cite{nkall}. The fields ${\hat \psi},~{\hat {\bar \psi}},~{\hat Q}$ are renormalized fields in in the renormalized QCD.

In section \ref{nqcd} we have shown that since the Yang-Mills theory is obtained by making analogy with the Maxwell theory by extending the U(1) gauge group to the SU(3) gauge group we find that as far as the boundary surface term at infinity is concerned the QCD predicts the same potential that is predicted by the classical Yang-Mills theory. This implies that if the boundary surface term at infinity is non-zero in the classical Yang-Mills theory then the corresponding boundary surface term at infinity is non-zero in QCD. The non-vanishing boundary surface term at infinity in the classical Yang-Mills theory gives non-vanishing angular momentum flux ${\vec J}_{\rm flux}$ as given by eq. (\ref{jfx}). Hence we find that the corresponding non-vanishing boundary surface term at infinity in QCD the gives non-vanishing angular momentum flux $<{\hat {\vec J}}_{\rm flux}>$ where
\bea
\frac{d<{\hat {\vec J}}_{\rm flux}>}{dt}\neq 0.
\label{jfxfn}
\eea
From eqs. (\ref{fne2x}) and (\ref{jfxfn}) we find that the gauge invariant total conserved angular momentum $<{\hat {\vec J}}_{total}>$ of all the quarks plus antiquarks plus gluons inside the proton is given by
\bea
<{\hat {\vec J}}_{total}>=\sum_{ q}<{\hat {\vec S}}_q>+\sum_{ q} <{\hat {\vec L}}_q>+ \sum_{\bar  q}<{\hat {\vec S}}_{\bar q}>+\sum_{\bar  q} <{\hat {\vec L}}_{\bar q}>+\sum_{g}<{\hat {\vec J}}_g>+<{\hat {\vec J}}_{\rm flux}>.\nonumber \\
\label{fne2y}
\eea
From eq. (\ref{fne2y}) we find
\bea
\frac{1}{2}=\sum_{  q}<{\hat { S}}_q>+\sum_{  q} <{\hat { L}}_q>+\sum_{\bar  q} <{\hat { S}}_{\bar q}>+\sum_{\bar  q} <{\hat { L}}_{\bar q}>+\sum_{g}<{\hat { J}}_g>+<{\hat { J}}_{\rm flux}>
\label{fnef}
\eea
which reproduces eq. (\ref{ini1}).

Hence from eq. (\ref{fnef}) we predict that the non-vanishing angular momentum flux contribution of the partons in QCD should be added to the spin and angular momentum of the partons to solve the proton spin crisis.

\section{ Absence of Non-Vanishing Boundary Surface Term in QCD in Study of Proton Spin Crisis in Literature is Not Justified }

In the section \ref{qcd} we saw that there exists non-vanishing boundary surface term $\frac{d<{\hat {\vec J}}_{\rm flux}>}{dt}$ in the angular momentum conservation equation in QCD in eq. (\ref{fne2x}) although the corresponding boundary surface term in QED is zero. This is because the potential energy at large distance $r$ in QED is a decreasing function of $r$ whereas the potential energy at large distance $r$ in QCD is an increasing function of $r$ due to confinement, a phenomenon which is absent in QED. Hence it is natural that this boundary surface term is zero in QED but is non-zero in QCD due to confinement.

One may argue that relativistic quark models with confinement, e.g. the MIT bag, correctly reproduce the angular sum-rule without any extra surface term contribution, see e.g. \cite{aw,alt,sd}, among others. One may also argue that there seems no need for the non-zero boundary surface term found in this paper in realistic QCD motivated models with confinement, which now do give a good description of the proton's spin structure once pion cloud effects are also included, see for example \cite{aw,alt,sd}. 

However, these arguments are not correct because QCD is the fundamental theory of the nature whereas the relativistic quark models with confinement, e.g. the MIT bag and the realistic QCD motivated models with confinement, see for example \cite{aw,alt,sd}, are not fundamental theory of the nature. Hence relativistic quark models with confinement, e.g. the MIT bag and the realistic QCD motivated models with confinement, see for example \cite{aw,alt,sd}, are not same as QCD. Since QCD is the fundamental theory of the nature describing the interaction between quarks and gluons, the spin of the proton from the spin and angular momentum of quarks and antiquarks and gluons inside the proton should be studied from the first principle by using the angular momentum conservation equation in QCD as given by eq. (\ref{fne2x}). Hence we find that the study of the proton spin crisis in the literature, see for example \cite{jaff,all1,aw,alt,sd} (among others), which did not include the non-vanishing boundary surface term $<{\hat {\vec J}}_{\rm  flux}>$ in QCD in eq. (\ref{fnef}) are not justified.

\section{ Angular Momentum Flux in QCD at Infinite Distance Is Not Same As The Anomalous Glue Contribution At Bjorken $x=0$ }

Note that if the boundary surface term is at finite distance then the boundary surface term is non-zero. We have found the non-vanishing boundary surface term $\frac{d<{\hat {\vec J}}_{\rm flux}>}{dt}$ at infinite distance in the angular momentum conservation equation in QCD in eq. (\ref{fne2x}) from the first principle.

It is useful to mention here the aspects of the anomalous glue contribution at Bjorken $x=0$ which arises from the gluon topology in the axial anomaly, see for example \cite{sd}. Note that the spin dependent parton distribution functions are not measured experimentally at high energy colliders at Bjorken $x=0$. Similarly the spin dependent parton distribution functions at high energy colliders at Bjorken $x=0$ are not calculated by using non-perturbative QCD. Hence in order to study the anomalous glue contribution at Bjorken $x=0$ one needs to study the  non-perturbative QCD at Bjorken $x=0$ which is not known. It is claimed in the literature that the anomalous contribution to $S_q$ associated with gluon topology and the axial anomaly at Bjorken $x=0$ cancels with a corresponding $x=0$ term in the orbital angular momentum since the total angular momentum is anomaly free \cite{sd,sd1}.

As far as the angular momentum flux in QCD and the anomalous glue contribution are concerned it is necessary to mention here that the anomalous glue contribution is not same as the angular momentum flux in QCD. This is because they have different field theoretical expressions. The expression of the angular momentum flux in eqs. (\ref{jfx}) and (\ref{jfxfn}) in QCD is obtained when the gluon lagrangian density is of the form \cite{nym}
\bea
{\cal L}_{gluon}(x) = -\frac{1}{4} F_{ \nu \lambda}^b(x)F^{\nu \lambda b}(x),~~~~~~~F_{\nu \lambda}^b(x)=\partial_\nu Q_\lambda^b(x)-\partial_\lambda Q_\nu^b(x)+gf^{bdc} Q_\nu^d(x) Q_\lambda^e(x)
\label{gll}
\eea
whereas the anomalous glue contribution is obtained from the anomalous gluon lagrangian density of the form \cite{shr}
\bea
{\cal L}_{\rm anomalous~gluon}(x) \propto g^2 \epsilon^{\nu \lambda \mu \delta} F_{ \nu \lambda}^b(x)F_{\mu \delta}^b(x).
\label{glt}
\eea
The best way to see that the angular momentum flux in QCD is different from the anomalous gluon contribution is to observe that there is no anomalous glue in the classical Yang-Mills theory but there is non-zero angular momentum flux in the classical Yang-Mills theory. Hence one finds that the angular momentum flux in QCD at infinite distance is not same as the anomalous glue contribution at Bjorken $x=0$. Therefore any conclusion drawn in the anomalous glue and axial anomaly scenario at Bjorken $x=0$ is not valid for the angular momentum flux in QCD at the infinite distance.

\section{Conclusions}
The proton spin crisis remains an unsolved problem in physics. In this paper we have found that due to the confinement of partons inside the hadron the angular momentum sum rule in QCD is violated. Hence we have found that the non-vanishing angular momentum flux contribution of the partons in QCD should be added to the spin and angular momentum of the partons to solve the proton spin crisis.

\end{document}